\documentstyle[12pt]{article}

\begin{document}

\sloppy

\begin{flushright}{UT-748}\end{flushright}

\vskip 1.5 truecm

\centerline{\large{\bf Large N Expansion and }}
\centerline{\large{\bf  Softly Broken Supersymmetry}}
\vskip .75 truecm
\centerline{\bf Tomohiro Matsuda
\footnote{matsuda@danjuro.phys.s.u-tokyo.ac.jp}}
\vskip .4 truecm
\centerline {\it Department of Physics, University of Tokyo}
\centerline {\it Bunkyo-ku, Tokyo 113,Japan}
\vskip 1. truecm

\makeatletter
\@addtoreset{equation}{section}
\def\theequation{\thesection.\arabic{equation}}
\makeatother

\vskip 1. truecm

\begin{abstract}
\hspace*{\parindent}
We examine the supersymmetric non-linear O(N) sigma model
with a soft breaking term.
In two dimensions, we found that
the mass difference between supersymmetric partner fields vanishes
accidentally. 
In three dimensions, the mass difference is observed but 
O(N) symmetry is always broken also in the strong coupling region.

\end{abstract}
\newpage

\section{Introduction}
\hspace*{\parindent}
Supersymmetric field theories have many attractive features.
For example, they may lead to the solution of the hierarchy
problem or the non-renormalizability of the quantum gravity.
While supersymmetry is theoretically attractive, it is not a
manifest symmetry of nature.
This necessitates the establishment of a realistic mechanism
of supersymmetry breaking for these theories.
In phenomenological models, we naively add soft breaking terms to the
supersymmetric models and break supersymmetry at tree level.

In this letter we first 
re-examine the supersymmetric non-linear O(N) sigma model
in two and three dimensions.
In two dimensions, both supersymmetry and O(N) symmetry
are not broken for any value of $g$.
In three dimensions, however, we can find two phases.
In the weak coupling phase, supersymmetry is not broken but
O(N) symmetry is broken.
In the strong coupling phase, both supersymmetry and O(N) symmetry
are preserved.
Next we examine the theory with a soft breaking term.
In two dimensions, we found that
the mass difference between supersymmetric partner fields
accidentally  vanishes but the supersymmetry is broken. 
In three dimensions, the mass difference is always observed and 
O(N) symmetry is always broken also in the strong coupling region.

\section{Large N expansion and softly broken supersymmetry}
\hspace*{\parindent}
The supersymmetric non-linear sigma model is usually defined by
the Lagrangian
\begin{equation}
  L=\frac{1}{2}\int{d}^{2}\theta\Phi_{j}D^{2}\Phi_{j}
\end{equation}
with the non-linear constraint
\begin{equation}
  \label{const2}
  \Phi_{j}\Phi_{j}=\frac{N}{g^{2}}.
\end{equation}
where the sum of the flavor index j runs from 1 to N.
The superfields $\Phi_{j}$ may be expanded out in components
\begin{equation}
  \Phi_{j}=n_{j}+\overline{\theta}\psi_{j}+\frac{1}{2}
  \overline{\theta}\theta{F}_{j}
\end{equation}
and the super covariant derivative is
\begin{equation}
  D=\frac{\partial}{\partial\theta}-i\overline{\theta}
  \not{\! \partial}.
\end{equation}
In order to express the constraint (\ref{const2}) as a $\delta$
function, we introduce a Lagrange multiplier superfield $\Sigma$.
\begin{equation}
  \Sigma=\sigma+\overline{\theta}\xi+\frac{1}{2}\overline{\theta}
  \theta\lambda
\end{equation}
We thus arrive at the manifestly supersymmetric action for the
supersymmetric sigma model\cite{alv}.
\begin{equation}
  \label{lag3}
  S=\int{d}^{D}xd^{2}\theta\left[\frac{1}{2}\Phi_{j}D^{2}\Phi_{j}
  +\frac{1}{2}\Sigma\left(\Phi_{j}\Phi_{j}-\frac{N}{g^{2}}
  \right)\right]
\end{equation}
In the component form, the Lagrangian from (\ref{lag3}) is
\begin{eqnarray}
  L&=&-\frac{1}{2}n_{j}\partial^{2}n_{j}+\frac{i}{2}
  \overline{\psi}_{j}\not{\! \partial}\psi_{j}+\frac{1}{2}
  F_{j}^{2}
  -\sigma{n}_{j}F_{j}-\frac{1}{2}\lambda{n}_{j}^{2}\nonumber\\
   &&+\frac{1}{2}\sigma\overline{\psi}_{j}\psi_{j}+\overline{\xi}
  \psi_{j}n_{j}+\frac{N}{2g^{2}}\lambda
\end{eqnarray}
We can see that $\lambda, \xi,$ and $\sigma$ are the respective
Lagrange multiplier for the constraints:
\begin{eqnarray}
  \label{what}
  n_{j}n_{j}&=&\frac{N}{g^{2}}\nonumber\\
  n_{j}\psi_{j}&=&0\nonumber\\
  n_{j}F_{j}&=&\frac{1}{2}\overline{\psi}_{j}\psi_{j}
\end{eqnarray}
The second and the third constraints of (\ref{what})
 are supersymmetric
transformations of the first.
We must not include kinetic terms for the field $\sigma$ and $\xi$
so as to keep these constraints manifest.
We can derive gap equations from 1-loop effective potential or
directly from eq.(\ref{what}) by using the tadpole method\cite{tad}.
These two approaches coincide to give the following equations.

(1) Scalar part\\
\begin{eqnarray}
  \label{1}
  n_{j}n_{j}|_{m_{n}^{2}=<\lambda>+<\sigma>^{2}}&=&
  N\int\frac{d^{D}p}{(2\pi)^{D}}
  \frac{1}{p^{2}+<\lambda>+<\sigma>^{2}}
  \nonumber\\
  &=&\frac{N}{g^{2}}
\end{eqnarray}

(2) Fermionic part\\
\begin{eqnarray}
  \label{2}
  \frac{1}{2}\overline{\psi}_{j}{\psi}_{j}|_{m_{\psi}=<\sigma>}
  &=&n_{j}F_{j}
\end{eqnarray}
This relation includes auxiliary field $F_{j}$, to be
eliminated by equation of motion.
After substituting $F_{j}$ by $\sigma{n}_{j}$, we obtain:
\begin{eqnarray}
  n_{j}F_{j}&=&\sigma n_{j}n_{j}
\end{eqnarray}
If we impose the O(N) symmetric constraint
$n^{2}=\frac{N}{g^{2}}$, we have
\begin{eqnarray}
  \frac{N}{g^{2}}\sigma&=&\frac{1}{2}
  \overline{\psi_{j}}\psi_{j}|_{
    m_{\psi}=<\sigma>}\nonumber\\
  \frac{N}{g^{2}}&=&\int\frac{dp^{D}}{(2\pi)^{D}}
  \frac{1}{p^{2}+<\sigma>^{2}}.
\end{eqnarray}

Let us examine these two equations in two and three dimensions.

For $D=2$ we obtain from scalar part:
\begin{eqnarray}
  \label{a}
  1&=&\frac{g^{2}}{4\pi}log\frac{\Lambda^{2}}{<\lambda>+<\sigma>^{2}}
  \nonumber\\
  m_{n}^{2}&=&
  <\lambda>+<\sigma>^{2}\nonumber\\
  &=&\Lambda^{2}exp\left(-\frac{4\pi}{g^{2}}\right)
\end{eqnarray}

And from fermionic part:
\begin{eqnarray}
  \label{2kai}
  m_{\psi}^{2}&=&
  <\sigma>^{2}\nonumber\\
  &=&\Lambda^{2}exp\left(-\frac{4\pi}{g^{2}}\right).
\end{eqnarray}
Substituting $<\sigma>$ in the scalar constraint (\ref{a}) with
(\ref{2kai}), we can find that $<\lambda>$ must vanish.
This means that  $\psi$ gains the same mass as $n$,
and simultaneously the supersymmetric order parameter $<\lambda>$
vanishes.
We can say that the supersymmetry is not broken in two dimensions
as is predicted by Witten\cite{witten}.

For D=3, the situation is slightly different.
We have a critical coupling constant $g^{2}_{cr}$ defined by:
\begin{equation}
  1=g^{2}_{cr}\int\frac{d^{3}p}{(2\pi)^{3}}\frac{1}{p^{2}}.
\end{equation}
If we take $g^{2}<g_{cr}^{2}$ something goes wrong with (\ref{1}).
It does not have any solution, so the constraint
$<\vec{n}^{2}>=\frac{N}{g^{2}}$
cannot be satisfied.
Of course, it is illusionary.
We should also consider the possibility of spontaneous breaking
of the O(N) symmetry.
In above discussions, we have implicitly assumed that the
vacuum expectation value of $\vec{n}$ would vanish.
Let us consider what may happen if $\vec{n}$ itself gets non-zero
vacuum expectation value.
Because of the O(N) symmetry, the vacuum expectation value of
$\vec{n}\equiv(n_{1},n_{2},...n_{N})$ may be written as
\begin{equation}
  <\vec{n}>=(0,0,...\sqrt{N}v/g).
\end{equation}
So that the constraint equation (\ref{1}) becomes
\begin{eqnarray}
  \label{gapv}
  n_{j}n_{j}|_{m_{n}^{2}=<\lambda>+<\sigma>^{2}}
  &=&N\left(\frac{v^{2}}{g^{2}}+\int\frac{d^{3}p}{(2\pi)^{3}}
  \frac{1}{p^{2}+<\lambda>+<\sigma>^{2}}\right)\nonumber\\
  &=&\frac{N}{g^{2}}.
\end{eqnarray}
Then we have an another critical coupling constant $g'_{cr}$:
\begin{equation}
  \label{v}
  \frac{1-v^{2}}{g_{cr}^{'2}}=\int\frac{d^{3}p}{(2\pi)^{3}}
  \frac{1}{p^{2}}
\end{equation}
If $g$ is smaller than $g_{cr}$, then $v$ grows. As a result,
the constraint equation has a solution in the weak coupling
 region($g'_{cr}\leq{g}\leq{g}_{cr}$)
 in a sense that not eq.(\ref{1}) but eq.(\ref{gapv})
is satisfied by some $<\lambda>+<\sigma>^{2}$.

Then what will happen if we include the fermionic part?
As far as $g\geq{g}_{cr}$, we have nothing to worry about.
In the strong coupling region, both supersymmetry and
the O(N) symmetry are preserved as we have explained in two dimensional
case.
However, in the weak coupling region, the situation is changed.
There is no non-trivial solution for the constraint (\ref{2}) and
there is no fermionic condensation that means 
no dynamical mass is generated for the fermion. 
It does not matter 
because we can set the supersymmetry breaking
order parameter $\lambda=0$ and then scalar field becomes massless
as well.
One may wonder why $\lambda=0$ is favorable, but we can easily find that
non-zero $\lambda$ can be related to the positive vacuum energy 
if we also consider the effective kinetic term for the auxiliary
superfield $\Sigma$.

So we can conclude:

(1) In two dimensions, both supersymmetry and the O(N) symmetry
are not broken.
This means that  $\lambda$  and $v$ remain zero for any
value of $g$.

(2) In three dimensions, both supersymmetry and the O(N) symmetry
are not broken (i.e., $\lambda$  and $v$ remain zero) in the strong
coupling region.
The O(N) symmetry can be broken in the weak coupling region, but
supersymmetry is kept unbroken in both phases.

Now let us extend the above analysis to
include a supersymmetry breaking mass term.
Here we consider:
\begin{equation}
  \label{soft}
  L_{break}=m_{s}^{2} n_{j}^{2}
\end{equation}
We can explicitly calculate the gap equation.
For the scalar part:
\begin{eqnarray}
  n_{j}n_{j}|_{m_{n}^{2}=<\lambda>+<\sigma>^{2}+m_{s}^{2}}&=&
  N\int\frac{d^{D}p}{(2\pi)^{D}}
  \frac{1}{p^{2}+<\lambda>+<\sigma>^{2}+m_{s}^{2}}
  \nonumber\\
  &=&\frac{N}{g^{2}}
\end{eqnarray}
The fermionic part is unchanged by the breaking term.
For $D=2$ we can solve this equation explicitly.
\begin{eqnarray}
  1&=&\frac{g^{2}}{4\pi}log\frac{\Lambda^{2}}
  {<\lambda>+<\sigma>^{2}+m_{s}^{2}}
  \nonumber\\
  m_{n}^{2}&=&
  <\lambda>+<\sigma>^{2}+m_{s}^{2}\nonumber\\
  &=&\Lambda^{2}exp\left(-\frac{4\pi}{g^{2}}\right)
\end{eqnarray}
$<\sigma>^{2}$ is determined by the fermionic part which is unchanged by 
the supersymmetry breaking term(\ref{soft}).
\begin{eqnarray}
  m_{\psi}^{2}&=&
  <\sigma>^{2}\nonumber\\
  &=&\Lambda^{2}exp\left(-\frac{4\pi}{g^{2}}\right).
\end{eqnarray}

These two equations suggest two consequences.
One is that the supersymmetry breaking parameter $\lambda$ gets non-zero
value:
\begin{equation}
  <\lambda>+m^{2}_{s}=0
\end{equation}
So the supersymmetry is broken.
The second is rather curious.
As we can see from explicit calculations,
 dynamically generated masses are unchanged so the mass degeneracy
is not removed.
This happens because the auxiliary field $\lambda$ has absorbed 
$m_{s}$ so that the two masses balance.

So we conclude that, if we believe the large N expansion, the
dynamical masses are unchanged while the supersymmetry breaking parameter
develops non-zero value .

The crucial point of our observation lies in the fact that
we can absorb the soft term by redefining a field.
The simplest and trivial example is the ordinary O(N) non-linear
sigma model with an explicit mass term.
This is written as:
\begin{equation}
  L=-\frac{1}{2}n_{j}\partial^{2}n_{j}
  -\frac{1}{2}\lambda(n_{j}^{2}-\frac{N}{g^{2}})
  -m^{2}n_{j}^{2}
\end{equation}
Does the explicit mass term changes the dynamical mass?
The answer is no.
This can easily be verified by redefining $\lambda$ as
$\lambda'=\lambda+m^{2}$.
Lagrangian is now:
\begin{equation}
  L=-\frac{1}{2}n_{j}\partial^{2}n_{j}-
  \frac{1}{2}\lambda'(n_{j}^{2}-\frac{N}{g^{2}})
  -\frac{N}{2g^{2}}m^{2}
\end{equation}
We can find that the mass term is absorbed in $\lambda$ and
only a constant is left.
Of course, this constant does not change the gap equation.

In three dimensions, however, it is not so simple.
Many fields and their equations form complex relations and determine 
their values.

Let us see more detail.
In three dimensions, we should slightly alter the above results.
As is discussed above, 
this model has a weak coupling region where no dynamical mass is produced
so  no balancing effect between superpartner masses
works in this region.
Setting $\lambda=0$, we find $m_{n}=m_{s}$ and $m_{\psi}=0$ when $g$ is
small.
This agrees with the naive expectation.
What will happen if we go into the strong coupling region
where the gap equation has non-trivial solution and
  the fermion becomes massive?
If there is no soft term, O(N) symmetry restoration occurs in this region.
But when $m_{s}$ is non-zero, $v$ must develop non-zero value
in order to compensate $m_{s}$ and 
satisfy the constraint equation(\ref{gapv}).
In this case, we can set $\lambda=0$ while $v$ becomes non-zero.

To summarize, after adding a breaking term, some
fields slide to compensate $m_{s}$ but the mechanism is not trivial.
Even in our simplest model, many complex relations determine their
values.

\section{Conclusion}
\hspace*{\parindent}

We examined the supersymmetric non-linear O(N) sigma model
 with a soft breaking term.
In two dimensions, we found that
the mass difference between supersymmetric partner fields vanishes
accidentally but the supersymmetry is broken. 
In three dimensions, the mass difference is always observed but 
O(N) symmetry is always broken even in the strong coupling region.

\section*{Acknowledgment}
\hspace*{\parindent}
We thank K.Fujikawa, T.Hotta and K.Tobe for many helpful discussions.


\begin{thebibliography}{1}

\bibitem{alv}
  O.Alvarez, Phys.Rev.D17(1978)1123\\
  T.Matsuda hep-ph/9605364

\bibitem{tad}
S.Weinberg Phys.Rev.D7(1973)2887\\
R.Miller, Phys.Lett.124B(1983)59, Nucl.Phys.B241(1984)535\\
T.Matsuda, J.Phys.A28(1995)3809

\bibitem{witten}
E.Witten, Nucl.Phys.B202(1982)253


\end{thebibliography}
\end{document}